\documentclass[a4paper,11pt]{article}
\usepackage{jheppub}
\usepackage[nameinlink, capitalize]{cleveref}
\usepackage{xspace}
\usepackage{cite}

\usepackage[dvipsnames]{xcolor}
\definecolor{EmeraldGreen}{HTML}{1ea78d}
\definecolor{EnglishRed}{HTML}{b02427}
\hypersetup{
	colorlinks=true,
	citecolor=EmeraldGreen,
	linkcolor={red!50!black},
	urlcolor=EmeraldGreen,
}

\preprint{MBI-ML-26-05, TIF-UNIMI-2026-10}

\newcommand{\hepml}{HEP--ML\xspace}

\title{\boldmath \huge The Living Guide of Machine Learning for Particle Physics}

\author[a]{Claudius Krause,}
\author[b]{Ramon Winterhalder,}
\author[c]{Matthew Feickert,}
\author[d,e]{Benjamin Nachman}

\affiliation[a]{Marietta Blau Institute for Particle Physics, Austrian Academy of Sciences\\
Dominikanerbastei 16, 1010 Wien, Austria}
\affiliation[b]{Università degli Studi di Milano \& INFN Sezione di Milano\\
Via Celoria 16, I-20133 Milano, Italy}
\affiliation[c]{University of Wisconsin--Madison\\
Madison, WI 53706, USA}
\affiliation[d]{Fundamental Physics Directorate, SLAC National Accelerator Laboratory\\
Menlo Park, CA 94025, USA}
\affiliation[e]{Department of Particle Physics and Astrophysics, Stanford University\\
Stanford, CA 94305, USA}

\emailAdd{claudius.krause@oeaw.ac.at}
\emailAdd{ramon.winterhalder@unimi.it}
\emailAdd{matthew.feickert@cern.ch}
\emailAdd{nachman@stanford.edu}

\abstract{
We started the Living Review of Machine Learning for Particle Physics (\hepml Living Review) in 2020 as a community-maintained, near-comprehensive bibliography of machine learning in particle physics.
The field was then growing faster than any single researcher could follow, finding the relevant papers was hard, and a structured, continuously updated reference paid off immediately.
Since then the literature has grown by more than an order of magnitude, the methods reach far beyond the classification and generation tasks of the early years, and the community has built its own ecosystem of topic-specific reviews, benchmark papers, and software frameworks.
The original model no longer serves this field well, and we can no longer sustain it.
We therefore change direction.
We freeze the Living Review as an archival reference covering the literature up to 1 June 2026, where it remains a stable record of the first phase of \hepml.
A new resource, the \emph{\hepml Living Guide}, replaces it.
It does not list everything.
It curates, it annotates, and it points readers to foundational and representative work, so that researchers can find their way into a mature and rapidly diversifying field.
In this article we explain why we make this change and how the new resource works.
}

\begin{document}
\maketitle

\clearpage
\section{Introduction}

\begin{figure}[b!]
    \centering
    \includegraphics[width=0.9\linewidth]{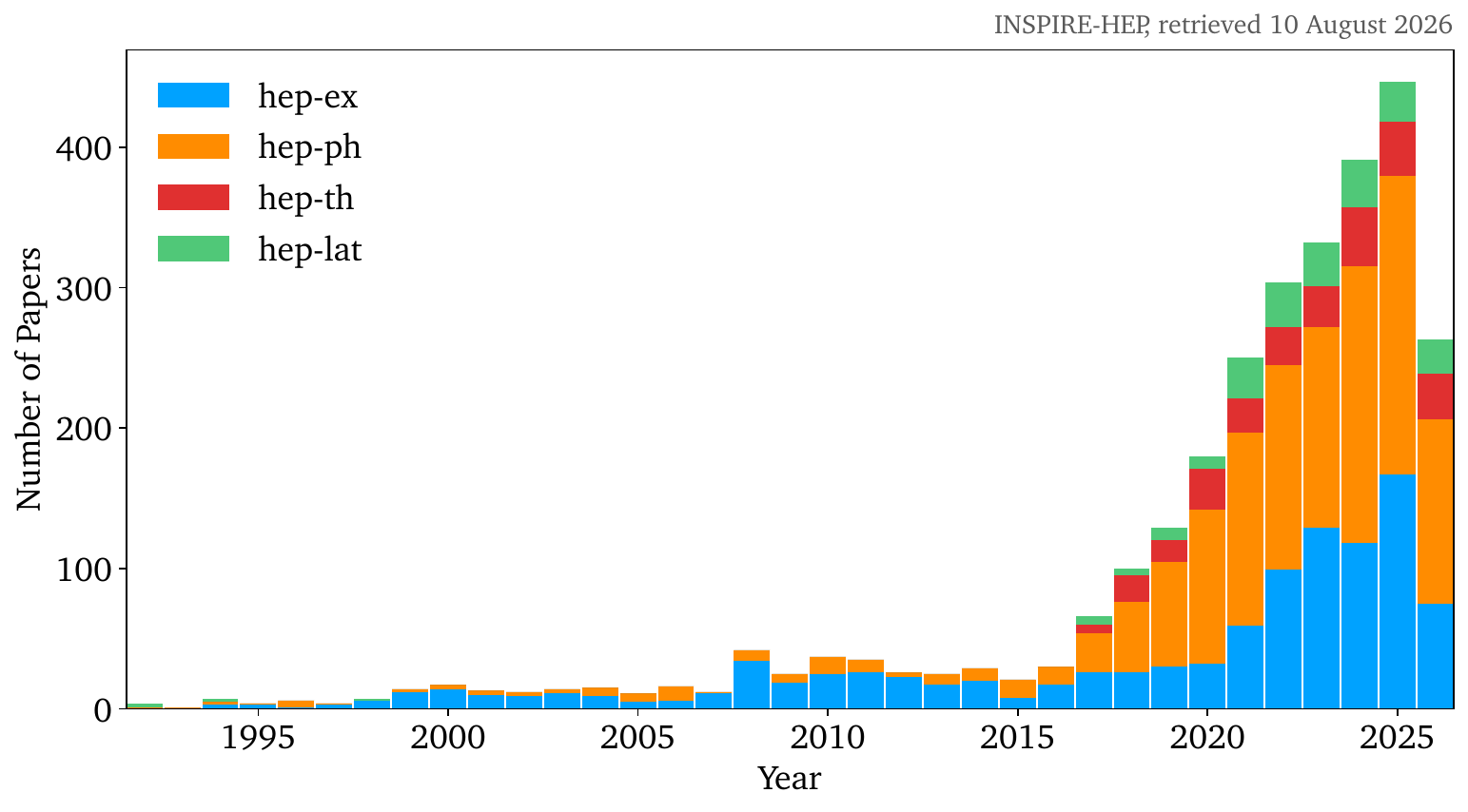}
    \caption{Number of papers per year on INSPIRE-HEP containing 
    the terms ``machine learning'', ``deep learning'', or ``neural network'', 
    broken down by HEP subcategory: phenomenology (\texttt{hep-ph}), 
    theory (\texttt{hep-th}), experiment (\texttt{hep-ex}), and lattice 
    (\texttt{hep-lat}). Data retrieved via the INSPIRE-HEP API.}
    \label{fig:trend_chart}
\end{figure}

Machine learning has become a central tool of modern particle physics.
It started as a niche activity around a handful of jet-tagging and event-classification problems.
Today it reaches across the whole field, from triggering and reconstruction, through simulation and fast emulation, unfolding and inference, phenomenology and formal theory, all the way to foundation models and agent-based approaches.
The publication rate follows this growth.
The \hepml Living Review lists more than 4\,000 papers and gains several hundred each year, with no sign of slowing down, see \cref{fig:trend_chart}.

We created the original Living Review of Machine Learning for Particle Physics~\cite{Feickert:2021ajf,HEP-ML-URL,HEP-ML-Zenodo} in 2020, when the literature was growing fast but a reader could still navigate it and the main problem was finding the relevant papers at all.
We chose the format deliberately.
A static review article would have gone out of date within weeks or months, so we built a continuously updated, community-contributed bibliography in a public version-controlled repository.
The review was ``living''.
This model worked well for its time.

Today the community faces a different challenge.
Finding papers is no longer the hard part, because arXiv alerts, INSPIRE queries, and modern search tools do that job well.
The hard part is finding a way in.
A newcomer needs to know how to enter a subfield, which papers laid its foundations, what the field has settled and what it has not, and how to connect work from communities that describe related ideas in different words.
A near-complete list of citations does not answer any of these questions, however well we organize it, and it can even hide the answers.
Neighboring areas have built their own community-curated resources, for machine learning in cosmology~\cite{MLCosmology-URL}, astronomy data science~\cite{AwesomeAstroData}, and HEP software~\cite{AwesomeHEP}.
They also let us define a narrower scope for our own resource, as we discuss in \cref{sec:scope}.
Together they show a broader shift, away from exhaustive bibliographies and toward curated entry points, software maps, and field guides.

We therefore change the model.
We freeze and archive the original \hepml Living Review~\cite{HEP-ML-Zenodo}, where it keeps its value as a bibliographic record of the \hepml literature until 1 June 2026.
The \hepml Living Guide takes its place and builds on curation, annotation, and community guidance instead of completeness.
Like its predecessor, the \hepml Living Guide (\url{https://iml-wg.github.io/HEPML-LivingGuide/}) belongs to the \hepml community, and anyone working in the field can write, revise, or correct it.
In \cref{sec:original} we describe the original design and what we learned from it, in \cref{sec:field} how the literature outgrew that model, and in \cref{sec:principles} how we
structure and maintain the new resource.

\section{The original \hepml Living Review}
\label{sec:original}

When we started the \hepml Living Review, machine learning in particle physics was expanding quickly, but it was still compact enough that near-comprehensive coverage was both feasible and useful.
A newcomer in 2020 could read through the full list and come away with a real overview of the landscape.
We used four broad categories.
Classification, regression, generative models, and inference covered most active research directions, and each of them held a manageable number of entries.

Two design choices proved especially valuable.
First, we kept the resource \emph{living}.
We made it citable and updated it continuously instead of freezing it at submission.
Second, we kept it \emph{community-driven}.
Contributing through a pull request on GitHub cost authors very little, so they added their own work, and the coverage spread across subfields and experiments instead of tracking what a small team happened to know.
In the end more than 40 different git users committed to the review.
The \hepml Living Review then played several roles at once.
It served as a shared bibliography, as a structured overview of the field, and as a lightweight onboarding resource for new researchers.

\subsection{The 2023 update}

\begin{figure}[ht!]
    \centering
    \framebox{\includegraphics[width=0.48\linewidth]{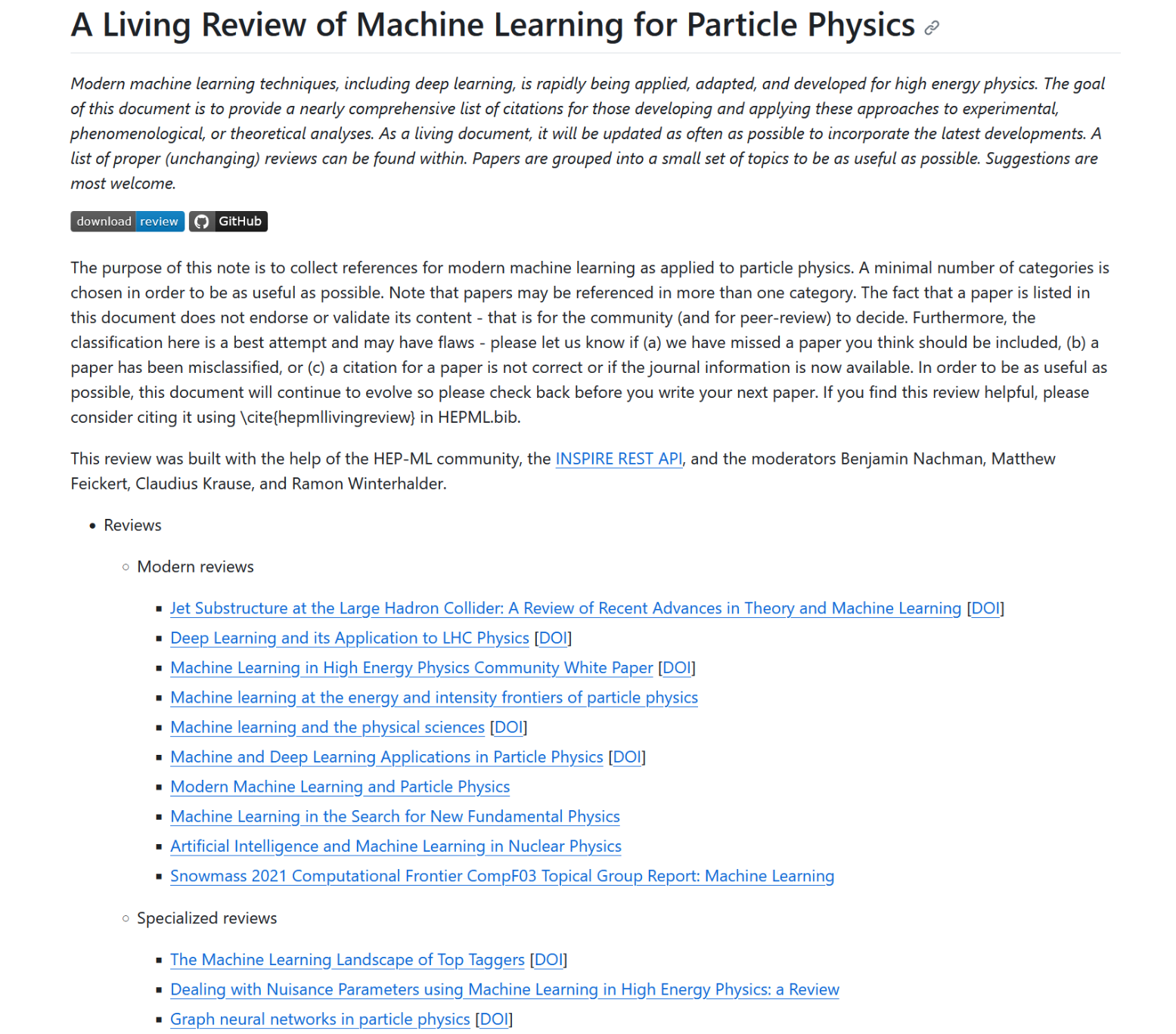}} \hfill 
    \framebox{\includegraphics[width=0.48\linewidth]{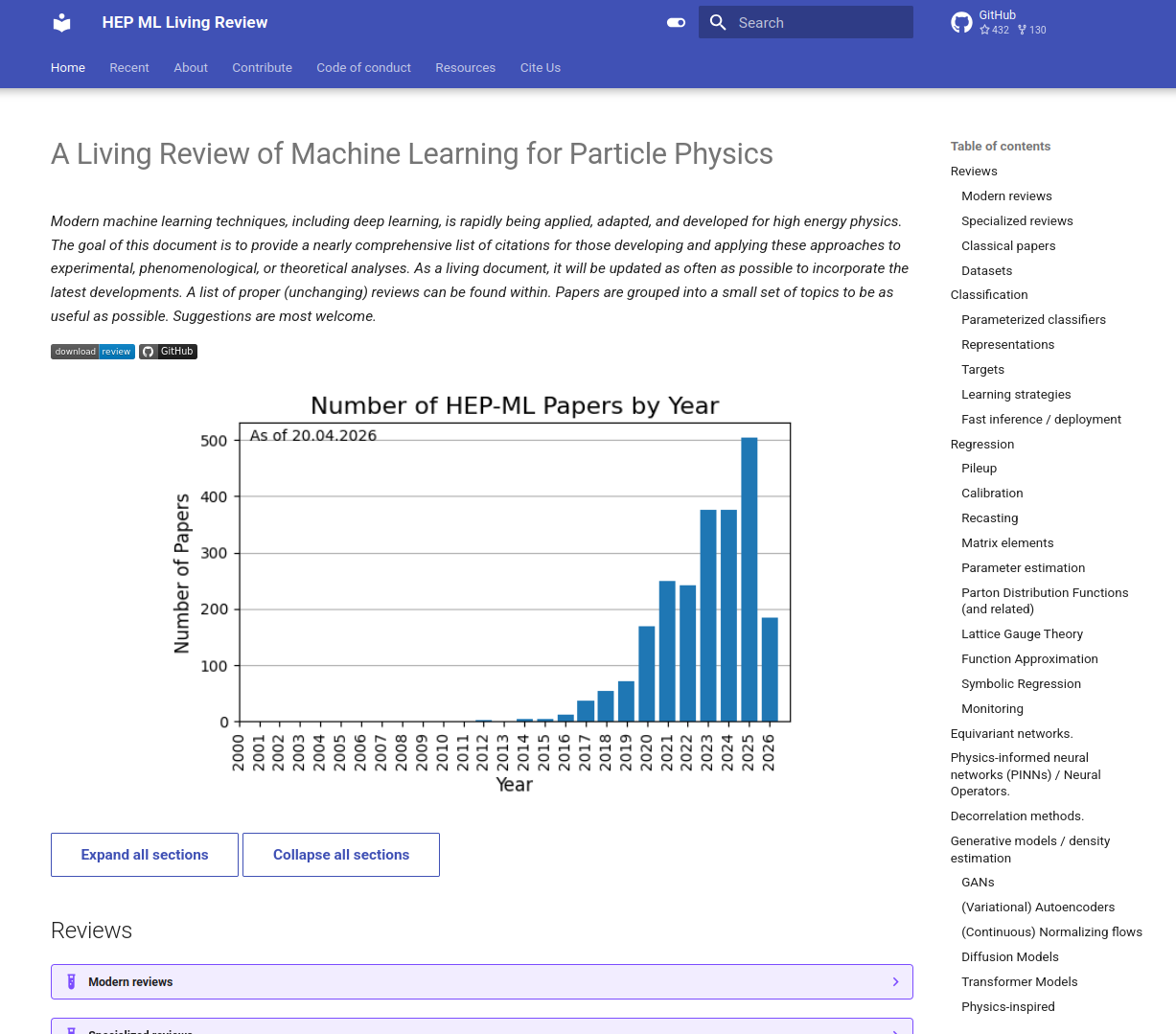}}
    \caption{The \hepml Living Review layout 2020--2023 (left) and 2023--2026 (right).}
    \label{fig:layout}
\end{figure}

In 2023 we rebuilt much of the web interface, see~\cref{fig:layout}.
We redesigned the visual layout and added a table of contents, a dedicated page for recent publications, and a search function.
These changes reflected an important diagnosis.
The main challenge was shifting from coverage to navigation.
The list had grown long enough that finding the right section took effort, and the taxonomy was showing its limits, because it had always mixed application targets with methodological categories.

At the same time we floated the idea~\cite{ML4Jets_talk} of restructuring the categories along two separate axes, HEP application and ML method.
The idea was sound, but we could not implement it without breaking backward compatibility, and reclassifying the entire review would have taken far too long.
This tension between the organizational demands of a maturing literature and the constraints of a flat, community-maintained list is one of the central lessons of the first phase.

\subsection{The evolution of the literature}
\label{sec:field}

Several concrete changes in the field motivate the transition to a new model.

\paragraph{Scale.}
The publication rate in HEP--ML has grown dramatically.
The Living Review added roughly 200 papers in its first year.
It now adds that many in a few months, see \cref{fig:trend_chart}.
Near-comprehensive manual curation at this rate demands more sustained effort than a small team of volunteers can give.
Unintended omissions become unavoidable.
The boundary of the field also blurred, once it had grown to include lattice QCD, neutrino, heavy-ion, nuclear, and astroparticle physics, along with a large overlap with formal theory.
A bibliography can absorb a vague boundary, because one extra entry costs almost nothing.
A curated guide cannot, because every choice about what to include says something about what the field is.

\paragraph{Breadth.}
Machine learning in particle physics no longer revolves only around classification and regression.
Researchers now work on anomaly detection,
simulation-based inference and unfolding, equivariant and geometric architectures, uncertainty quantification and calibration, differentiable programming and end-to-end optimization, hardware-aware deployment and triggering, foundation models trained on collider or detector data, and formal connections between ML theory and physics.
Many of these areas have grown their own review papers, benchmark datasets, community workshops, and dedicated community resources~\cite{SBI-URL1,SBI-URL2,ml_accel_review_2025}.
The field has spawned subfields.

\paragraph{Maturity.}
The community has meanwhile developed its own structure for organizing knowledge.
It has produced lecture notes~\cite{Plehn:2022ftl,Halverson:2024hax}, the \textit{Les Houches} guide to reusable ML models~\cite{Araz:2023mda}, summary papers of community challenges~\cite{Kasieczka:2019dbj,Kasieczka:2021xcg,Aarrestad:2021oeb,Krause:2024avx}, dedicated Snowmass reports~\cite{Shanahan:2022ifi,Butter:2022rso}, and a growing number of topic-specific reviews~\cite{Huetsch:2024quz,Aarrestad:2026xrs}.
The \hepml Living Review no longer has to serve as the only entry point into the literature.
It can do something more selective and more useful instead, and orient readers rather than enumerate papers.

\subsection{Lessons learned}
\label{sec:lessons}

\paragraph{What worked.}
The community contribution model worked.
A public GitHub repository with a low-friction pull-request workflow produced broader coverage than any small team could have achieved alone.
The resource became a standard citation in reviews and theses, which shows that the community adopted it, even though the credit did not always find its way back to the authors.
An INSPIRE-HEP fulltext search returns 81 entries that cite only the URL~\cite{HEP-ML-URL}, which adds nothing to the authors' citation count, 217 that cite only the arXiv article~\cite{Feickert:2021ajf}, which is frozen in time and not living, and 20 that cite both~\cite{HEP-ML-URL,Feickert:2021ajf}.
The Review was also a single rolling document.
Apart from the final archival snapshot it carried no tagged releases.
A citation could therefore point at the 2021 article or at the URL, but never at a particular state of the bibliography, and never at the people who produced that state.
Neither target reaches the more than 40 people who contributed to it.
Nevertheless, a citable and continuously updated living document was the right answer to a fast-moving field.

\paragraph{What became difficult.}
The flat taxonomy did not age well.
As the literature grew, the categories became too broad and too narrow at the same time.
They collected hundreds of papers with little in common, while still missing cross-cutting work that spanned several methods and applications.
We also missed more papers, never on purpose, as the volume and the breadth of the field kept growing.
Mixing methodological and application-oriented categories made it hard to answer the questions researchers really ask.
What is the best approach for unfolding?
Which architectures work for fast simulation?
Which papers established the theoretical foundations of simulation-based inference?

A comprehensive list answers the question ``what exists?'' but not ``what matters, and why?''
In a mature field the second question is the more valuable one.

\paragraph{What users need now.}
Conversations with researchers at all career stages show a consistent pattern.
Students and postdocs entering a new subfield do not want a list of every paper on the topic.
They want three to five papers to read first, two or three tutorials or reviews, and a map of how the subfield connects to its neighbors.
Experienced researchers moving between subfields ask for the same kind of entry point.
A comprehensive bibliography serves neither group as well as an annotated and curated resource does.

\section{The new \hepml Living Guide}
\label{sec:principles}

The \hepml Living Guide is built on the following commitments.

\begin{enumerate}
    \item \textbf{No claim of completeness.}
    We make no attempt to list all relevant papers.
    This is not a limitation we apologize for.
    It is the point.
    Completeness made sense as a goal in 2020, and it no longer does in 2026.

    \item \textbf{Curation through community interest.}
    The resource selects itself, because contributions follow real community investment in a topic.
    Three explicit criteria guide the selection.
    They are foundational importance, methodological clarity, and practical use for researchers entering the area.
    The community remains the driving force.
    A paper enters the Guide because researchers who work in the area consider it worth including and because it meets these criteria.
    We do not intend this as an additional layer of peer review.

    \item \textbf{Annotation is required.}
    Every curated paper or cluster of papers carries a short explanation of why we include it, what it establishes, and how it relates to neighboring work.
    A list of links without context is a search result, not a review.

    \item \textbf{Complement INSPIRE-HEP and arXiv, do not compete with them.}
    Existing infrastructure already handles comprehensive bibliographic search well.
    The \hepml Living Guide adds what those tools leave out, namely structured context and explicit guidance on where to start.

    \item \textbf{Sustainability by design.}
    Named community members write sections as one-time contributions, and we timestamp them so readers know how old they are.
    A section stays as it is until a new contribution updates it.
    We tag periodic releases of the Guide, so a citation points at a fixed state and at the people who wrote it.
    This keeps the maintenance burden light and the quality bar clear.
\end{enumerate}

\subsection{Scope}
\label{sec:scope}

The \hepml Living Guide covers machine learning for \emph{particle physics}, and we state that boundary explicitly, because curation needs one.
We include work that applies machine learning to, or develops it for, collider physics and phenomenology, formal and theoretical particle physics, lattice field theory, and neutrino physics.
Work belongs in the Guide when either the physics problem or the methodological development is specific to this domain.

We deliberately leave out several adjacent areas that the original Living Review had begun to absorb.
These are nuclear and heavy-ion physics, astroparticle physics, cosmology and astronomical data science, machine learning for accelerator design and operation, and generic machine-learning methodology without particle-physics content.
We narrow the scope on purpose relative to the archived Living Review, whose title still referred to particle \emph{and nuclear} physics.
A curated guide has to answer for what it recommends, and we cannot answer across that whole span.
This says nothing about the value of that work.
Several of these areas already run their own community-curated resources~\cite{MLCosmology-URL,AwesomeAstroData,ml_accel_review_2025}, and a guide that claimed to orient newcomers across all of them would do a bad job of it.
Where the boundary really is porous, we link to those resources instead of duplicating them.

\subsection{Structure of the \hepml Living Guide}
\label{sec:structure}

\begin{figure}[t!]
    \centering
    \framebox{\includegraphics[width=0.48\linewidth]{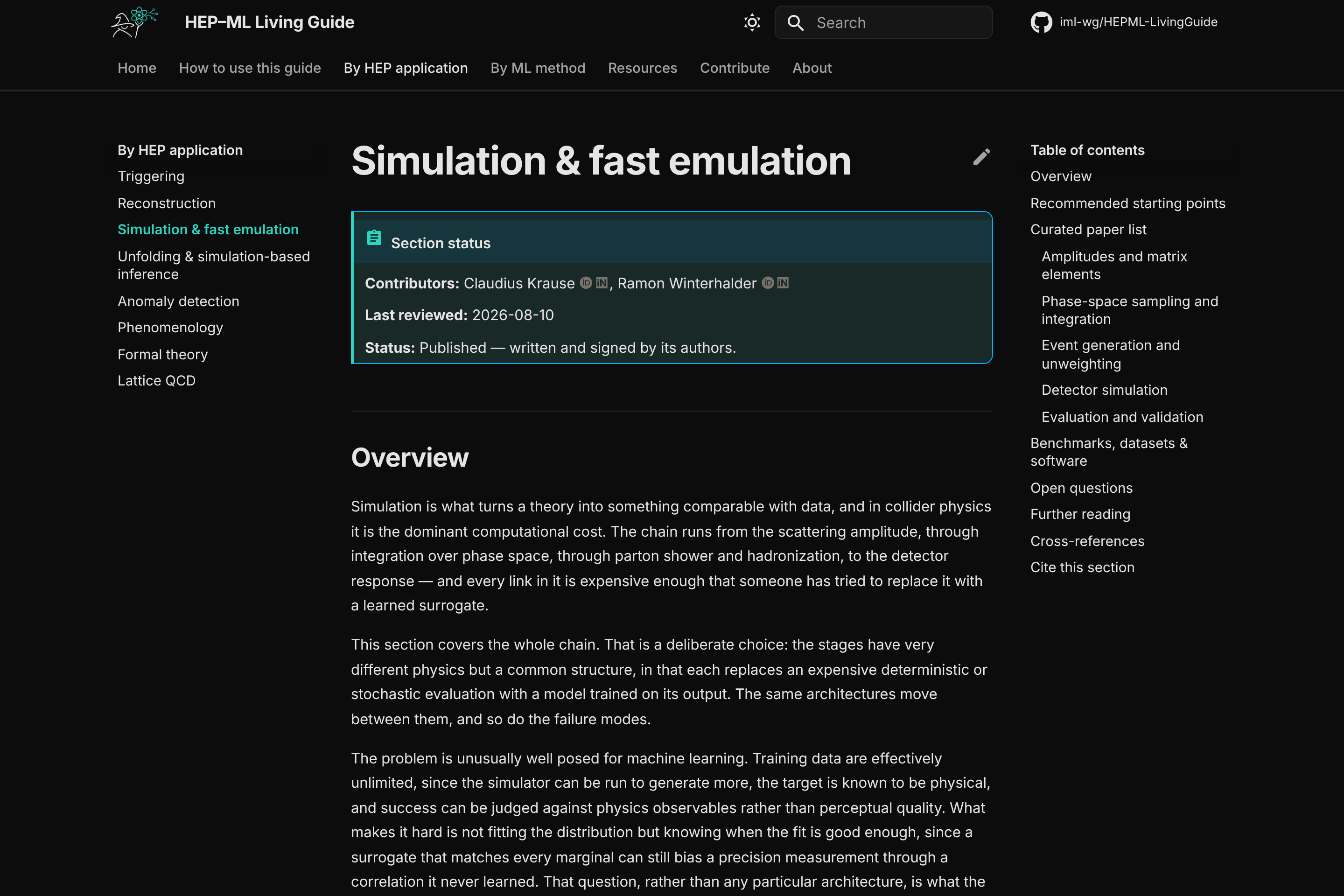}} \hfill
     \framebox{\includegraphics[width=0.48\linewidth]{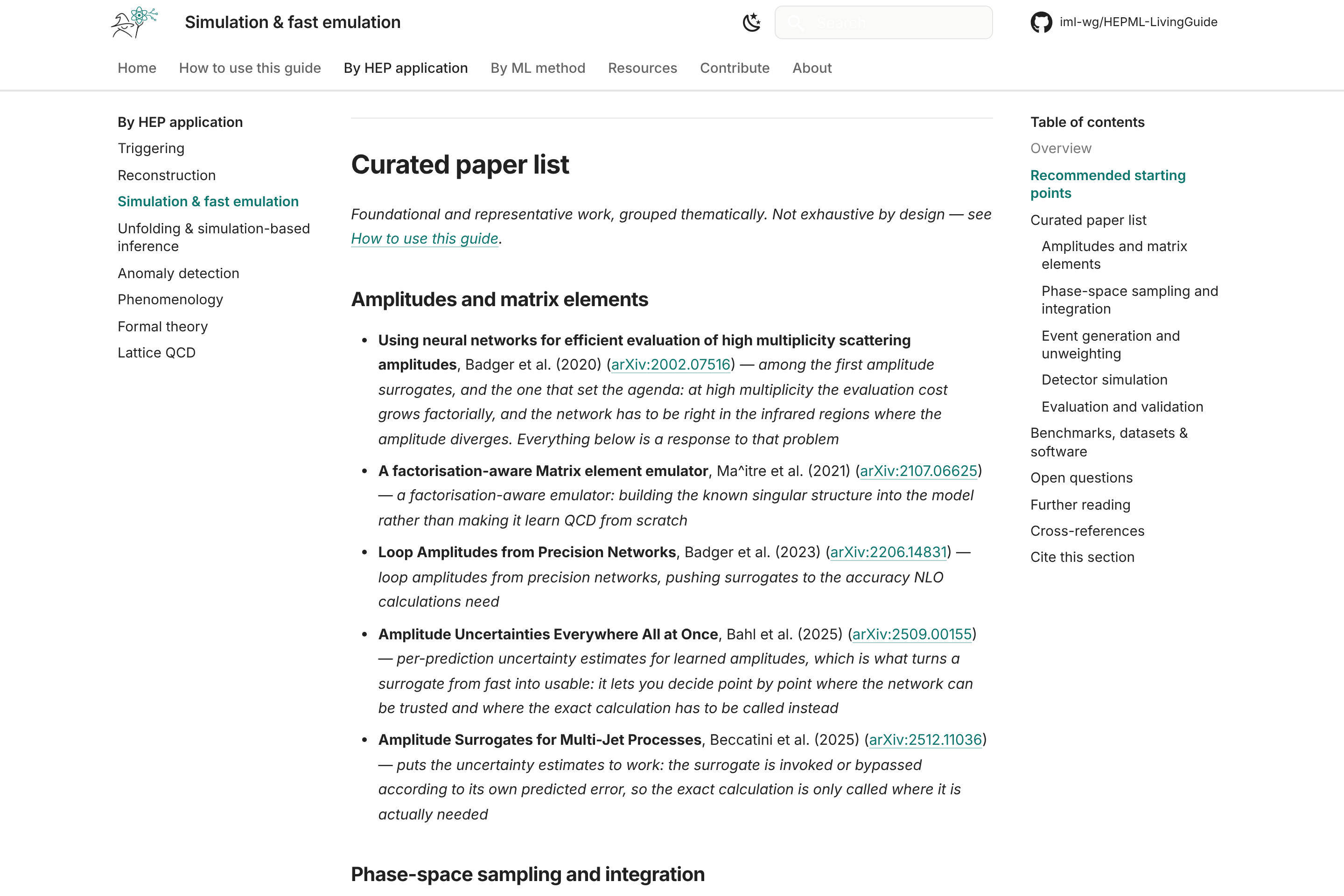}}
    \caption{The \hepml Living Guide. Left: a section header in dark mode, showing the named contributors, the date of last review, and the status of the section.
    Right: part of the curated list in light mode, with every entry annotated and
    grouped thematically, where the table of contents shows the structure each section follows.}
    \label{fig:new_layout}
\end{figure}

We organize the new resource as a curated field guide.
It splits into topical sections that follow the major research directions in HEP--ML within the scope of \cref{sec:scope}.
The taxonomy runs along two independent axes, \emph{HEP application} and \emph{ML method}, and we cross-link between them.
Anyone interested in fast calorimeter simulation goes straight to that application section.
Anyone who wants to know where the field uses diffusion models enters through the method axis and finds pointers indexed by application.

\subsubsection*{Section structure}

Each section contains the following elements:

\begin{itemize}
    \item \textbf{Overview.} A brief description of the problem, why it matters in the HEP context, and where it sits relative to adjacent areas (one to two paragraphs).
    \item \textbf{Recommended starting points.} A small set (typically three to six) of reviews, tutorials, or lecture notes, each with a one-sentence annotation on what it covers and whom it suits.
    \item \textbf{Curated paper list.} Foundational and representative papers, grouped thematically, each with a brief annotation. Not exhaustive by design.
    \item \textbf{Benchmarks, datasets, and software.} Key references for reproducibility and comparison, where they exist.
\end{itemize}

\subsubsection*{Contribution and maintenance}

Sections carry names.
A researcher or a small group with expertise in an area writes the first version, and we timestamp it and publish it under their name.
We keep the pull-request model for later additions and corrections, so anyone can still suggest a paper or flag an error, but we check every pull request against the stated criteria before we merge it.
We do not expect this to trigger controversial discussions that would call for extra layers of moderation.
The resource coordinators handle this review, which keeps the governance simple.
Because contributions select themselves, the active areas of the community attract the most frequent updates, and nobody has to coordinate centrally which sections need refreshing.
Researchers who write or substantially update a section appear by name, which gives them a concrete and citable record of their contribution to the community resource.
We therefore version the Guide.
At regular intervals we tag a release and archive it on Zenodo, where it gets its own DOI.
The author list of that release names everyone who wrote or substantially revised a section since the previous tag.
A reader then cites the release they used, and the citation reaches the people who wrote it.
This is the part of the old model we most wanted to repair.

\subsubsection*{The archived \hepml Living Review}

We preserve the original \hepml Living Review in its current state as a static archive.
Readers can still cite it through the original arXiv article~\cite{Feickert:2021ajf} and through a dedicated Zenodo record~\cite{HEP-ML-Zenodo} that captures the final archived state, and we keep the GitHub repository in read-only mode.
The archive remains a historical bibliographic record of the HEP--ML literature up to 1 June 2026, and it keeps its value for exactly that purpose.
Anyone who needs a comprehensive list of papers up to that date can go there directly and cite the snapshot as version \texttt{v2026.06.01}.
The \hepml Living Guide succeeds it.
It does not overwrite the archive, and we designed it for what the community needs now rather than for what it needed in 2020.

\section{Conclusions}

The Living Review of Machine Learning for Particle Physics was the right resource for 2020.
Its community-driven, near-comprehensive model solved a real problem in a fast-growing field, namely how to find the relevant papers at all.
The growth it helped document has now made that model obsolete.
The field has become too large and too mature for a near-complete manual bibliography to stay sustainable or particularly useful.

The \hepml Living Guide answers a different question.
It asks how to make the literature navigable, not how to record it.
We therefore curate instead of accumulate, we annotate instead of list, and we let community interest set the coverage.
We keep the pull-request model, but we ask contributors for a little more.
A sentence of context beats a bare link, and the resource becomes more useful for it.

We invite the community to contribute to the new resource, by writing sections, by suggesting additions, and by telling us what we got wrong or left out.
The resource stays living.
What changes is what ``living'' means in practice.

\subsection*{Acknowledgments}

We thank the entire \hepml community, who contributed to the \hepml Living Review over the past years and who now carry the Living Guide.
CK thanks the Università degli Studi di Milano, where parts of this document were written, for its hospitality.
MF is supported by the US National Science Foundation (NSF) under Cooperative Agreement PHY-2323298.
BN is supported by the US Department of Energy (DOE) under contract DE-AC02-76SF00515.

\clearpage
\bibliographystyle{tepml}
\bibliography{refs.bib}

\end{document}